\documentclass[epj-spec]{svjour}
\usepackage{epsf,epsfig}

\def\xb{\overline{x}}

\def\als{\alpha_s}
\def\vk{{\bf k}_{\perp}}

\def\gev{\,{\rm GeV}}
\begin{document}
\title{Diffractive vector meson electroproduction at small
Bjorken $x$ within GPD approach}
\author{S.V.Goloskokov \thanks{\email{goloskkv@theor.jinr.ru}} }
\institute{ Bogoliubov Laboratory of Theoretical  Physics,
  Joint Institute for Nuclear Research, Dubna, Russia}
\abstract{We study light vector meson electroproduction at small
$x$ within the generalized parton distributions (GPDs) model. The
modified perturbative approach is used, where the quark transverse
degrees of freedom in the vector meson wave function and hard
subprocess are considered. Our results on the cross section and
spin observables are in good agreement with experiment.
} 
\maketitle

 \section{Introduction}
Study of GPDs can give  extensive information on the hadron
structure. Vector meson leptoproduction at high energies
\cite{gk05,gk06,gk07t} is one of the essential processes where the
GPDs can be investigated. At small $x$-Bjorken  the leading twist
amplitude with longitudinally polarized photon and vector meson
(LL amplitude) at large photon virtualities $Q^2$ factorizes
\cite{fact} into a hard meson leptoproduction off partons and
GPDs. The transition amplitude for transversally polarized photons
$\gamma _{\perp }^{\ast }\rightarrow V_{\perp }$ (TT amplitude) is
suppressed as a power of $1/Q$ and its factorization is
problematic. The generally used collinear approximation  leads to
difficulties in studying  these reactions.  Really the
longitudinal cross section calculated in the collinear approach
\cite{mpw} exceeds the data by an order of magnitude. The TT
amplitude exhibits  infrared singularities in the collinear
approach \cite{mp} which is a reason of the factorization
breakdown.

 These problems can be solved in handbag model
\cite{gk05,gk06}  based on the modified perturbative approach
(MPA) \cite{sterman} which includes the quark transverse degrees
of freedom accompanied by the Sudakov suppressions. They suppress
the contribution from the end-point region and the LL cross
section are close to the experiment. The transverse quark momentum
regularizes the end-point singularities in the TT amplitudes and
it can be calculated in the model. Thus, our model provides an
access to study GPDs in the polarized vector meson production.

In our previous calculations \cite{gk05} we analyzed the low $x
\leq 0.01$ region where the gluon contribution had a predominant
role. In this report, we extend our analysis to moderate $x \sim
0.2$ \cite{gk06}. Within the MPA we calculate the LL and TT
 amplitudes, and afterwards the cross sections and the spin
observables in the light vector meson leptoproduction. Our results
are in reasonable agreement  with high energy  HERA experiments
\cite{h1,zeus} and low energy HERMES \cite{hermes} and E665
\cite{e665} data for electroproduced $\rho$ and $\phi$ mesons at
small $x$ \cite{gk05,gk06}.

\section{Leptoproduction of  Vector Mesons in the GPD approach}

The model is based on the handbag approach where the $\gamma^{\ast
} p\rightarrow V p$ amplitude  factorizes into hard partonic
subprocess and GPDs. In the region of small $x \leq 0.01$ the
gluon GPD  is a dominant contribution \cite{gk05}. At larger $x
\sim 0.2$ the quark contributions are essential \cite{gk06}.  For
small $t$ the amplitude of the vector meson production off the
proton with positive helicity  reads as a convolution of the hard
subprocess amplitude ${\cal H}^V$ and  GPDs
$H^i\,(\widetilde{H}^i)$:
\begin{eqnarray}\label{amptt-nf-ji}
  {\cal M}^V_{\mu'+,\mu +} &=& \frac{e}{2}\, {\cal
  C}^{V}\, \sum_{\lambda}
         \int d\xb
        {\cal H}^{Vi}_{\mu'\lambda,\mu \lambda}
                                   H^i(\xb,\xi,t) ,
\end{eqnarray}
where  $i$ denotes the gluon and quark contribution,
 $\mu$ ($\mu'$) is the helicity of the photon (meson), $\xb$
 is the momentum fraction of the
parton with helicity $\lambda$, and the skewness $\xi$ is related
to Bjorken-$x$ by $\xi\simeq x/2$. The flavor factors are
$C^{\rho}=1/\sqrt{2}$ and ${ C}^{\phi}=-1/3$. The polarized GPDs
$\tilde H^i$ are unimportant in the analysis of the cross section
because at small $x$ they are much smaller with respect to the
unpolarized GPDs $H^i$.

The hard part is calculated   using the $k$- dependent wave
function \cite{koerner} that contains the leading and higher twist
terms   describing the longitudinally and transversally polarized
vector mesons, respectively.
 The subprocess amplitude is calculated within the MPA
\cite{sterman}. The  amplitude ${\cal H}^V$ is represented as  the
contraction of the hard
  part $F$, which is calculated perturbatively, and the
non-perturbative meson  wave function $ \phi_V$
\begin{equation}\label{hsaml}
  {\cal H}^V_{\mu'+,\mu +}\,=
\,\frac{2\pi \als(\mu_R)}
           {\sqrt{2N_c}} \,\int_0^1 d\tau\,\int \frac{d^{\,2} \vk}{16\pi^3}
            \phi_{V}(\tau,k^2_\perp)\;
                  F_{\mu^\prime\mu}^\pm .
\end{equation}
  The wave function is chosen  in the Gaussian form
\begin{equation}\label{wave-l}
  \phi_V(\vk,\tau)\,=\, 8\pi^2\sqrt{2N_c}\, f_V a^2_V
       \, \exp{\left[-a^2_V\, \frac{\vk^{\,2}}{\tau\bar{\tau}}\right]}\,.
\end{equation}
 Here $\bar{\tau}=1-\tau$, $f_V$ is the
decay coupling constant and the $a_V$ parameter determines the
value of average transverse momentum of the quark in the vector
meson. The values of $f_V, a_V$ are  different for longitudinal
and transverse polarization of the meson.

In the hard part $F$ we keep the $k^2_\perp$ terms in the
denominators of LL and TT transitions and in the numerator of the
TT amplitude. The gluonic corrections are treated in the form of
the Sudakov factors which additionally suppress the end-point
integration regions.

To estimate  GPDs, we use the double distribution representation
\cite{mus99} with the double distribution function
\begin{equation}\label{ddf}
f_i(\beta,\alpha,t)= h_i(\beta,t)\,
                   \frac{\Gamma(2n_i+2)}{2^{2n_i+1}\,\Gamma^2(n_i+1)}
                   \,\frac{[(1-|\beta|)^2-\alpha^2]^{n_i}}
                          {(1-|\beta|)^{2n_i+1}}\,.
                          \end{equation}
The powers $n_i$ (i= gluon, sea, valence contributions) and the
functions $h_i(\beta,t)$ which are connected with parton
distributions are determined by
\begin{eqnarray}\label{pdf}
& h_g(\beta,0)=|\beta|g(|\beta|),& n_g=2;\nonumber\\
& h_{sea}^q(\beta,0)=q_{sea}(|\beta|) \mbox{sign}(\beta), & n_{sea}=2;\nonumber\\
& h_{val}^q(\beta,0)=q_{val}(\beta) \Theta(\beta), & n_{val}=1.
\end{eqnarray}
Here $g$ and $q$ are the ordinary gluon and quark PDF. For the
parton distribution the simple Regge anzats at small momentum
transfer is used
\begin{equation}\label{gd}
h_i(\beta,t)= e^{b_i t}\beta^{-(\delta_i(Q^2)+\alpha'_i
t)}\,(1-\beta)^{2 n_i+1}\sum_{j=0}^3\,c_i^j\, \beta^{j/2}.
\end{equation}
The  $\delta_i(Q^2)$ is connected  \cite{gk06} with the
corresponding intercept $\alpha_i(0)$ of the Regge trajectory
$\alpha_i=\alpha_i(0)+\alpha'_i t$. The parameters $c_i^j$ in
(\ref{gd}) and there evolution  are chosen by comparison with the
CTEQ6M PDFs
 \cite{CTEQ}.

The GPDs are related with PDFs through the double distribution
form
\begin{equation}
  H_i(\xb,\xi,t) =  \int_{-1}
     ^{1}\, d\beta \int_{-1+|\beta|}
     ^{1-|\beta|}\, d\alpha \delta(\beta+ \xi \, \alpha - \xb)
\, f_i(\beta,\alpha,t).
\end{equation}

\begin{figure}[h!]
\begin{center}
\begin{tabular}{ccc}
\mbox{\epsfig{figure=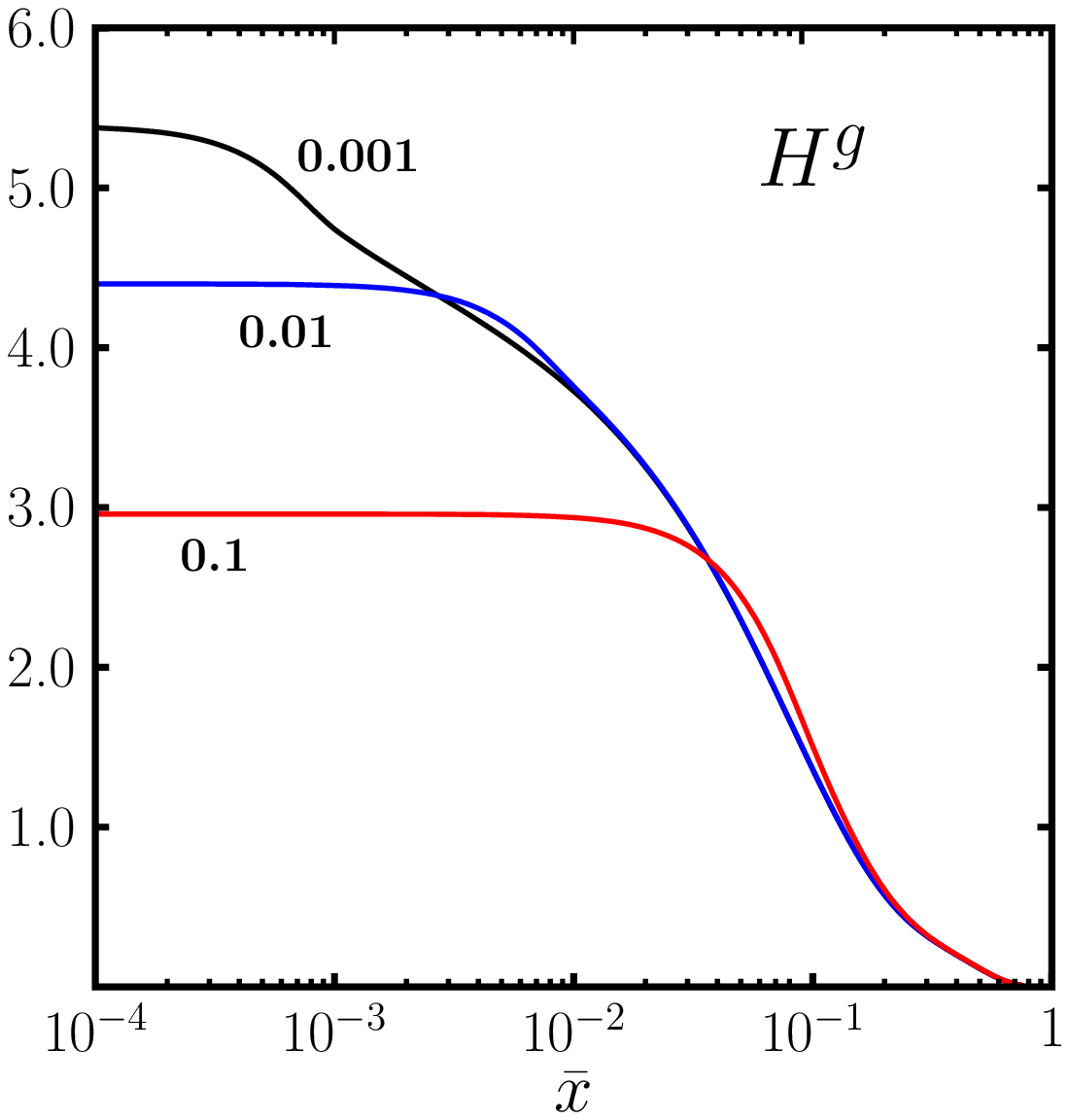,width=4.5cm,height=4.2cm}}&
\mbox{\epsfig{figure=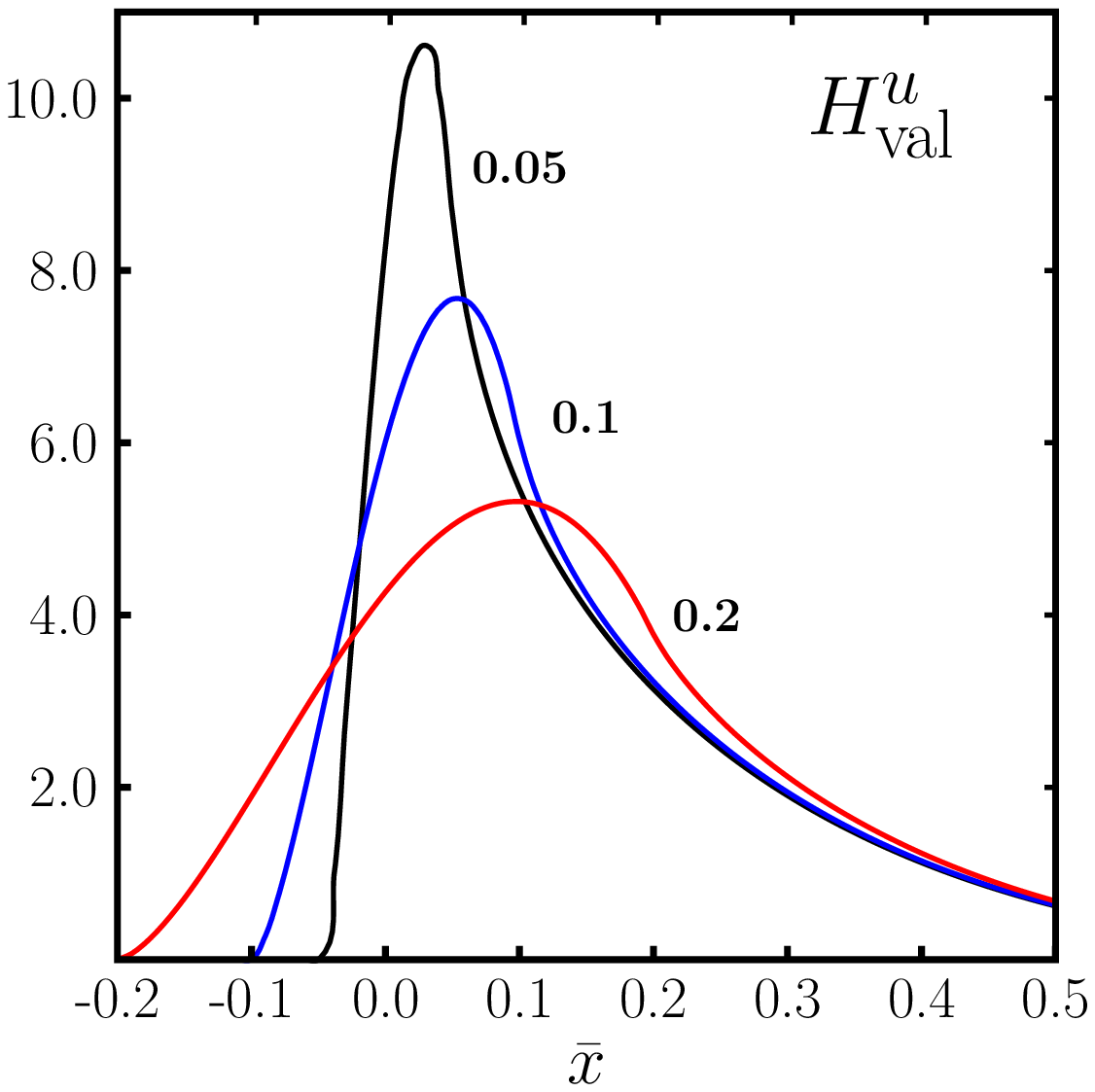,width=4.5cm,height=4.2cm}}&
\mbox{\epsfig{figure=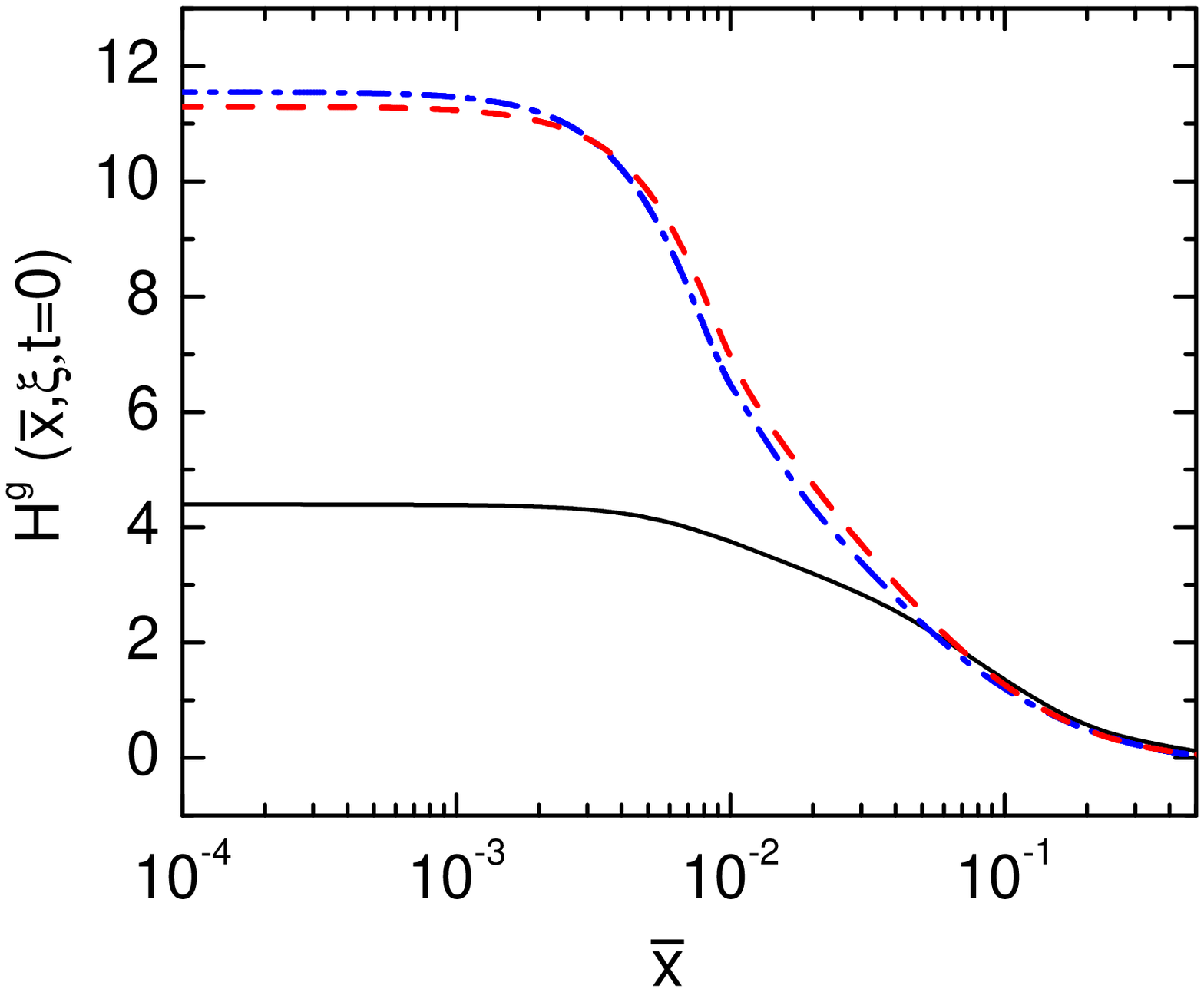,width=4.3cm,height=4.2cm}}\\
{\bf(a)}& {\bf(b)}& {\bf(c)}
\end{tabular}
\end{center}
\caption{ GPDs {\bf(a)} $H^g$, {\bf(b)} $H^u_{val}$ for some
values of skewness via $x$. GPDs are shown at $t =0$ and scale
$Q^2=4\mbox{GeV}^2$, {\bf(c)} Evolution of gluon GPD for $\xi=
0.01$. Full line- $Q^2=4\mbox{GeV}^2$; $Q^2=40\mbox{GeV}^2$:
dashed-dotted line- evolution code \cite{vinn}, dashed line- our
approximation.}
\end{figure}
The model results for the gluon and valence quark GPDs for  the
three $\xi$ values are shown in Fig. 1. In our model, the GPD
evolution is determined by the evolution of PDFs in
(\ref{ddf},\ref{gd}). To show that we reproduce the $Q^2$
dependence of GPDs correctly,  Fig. 1c represents the results of
our model and the GPDs evolution determined by evolution code
\cite{vinn}. Both results  coincide at $Q^2=4 \mbox{GeV}^2$ and
are close to each other at $Q^2=40 \mbox{GeV}^2$.

\section{Cross section and spin observables}
In this section we study the longitudinal cross section of the
vector meson production in the energy range $5 \mbox{GeV}< W< 75
\mbox{GeV}$ and the spin observables at HERA energies. The $t$-
dependence of the amplitudes is important in analyses of
 experimental data integrated over $t$. In our model, the diffraction peak
slopes  for LL and TT transitions are closed to each other:
$B_{LL}\sim B_{TT}$ and are determined from (\ref{gd})
\begin{equation}\label{bg}
B_i=2 \,b_i+2 \alpha_i'\ln\frac{W^2+Q^2}{Q^2+m_v^2}.
\end{equation}
For gluon and sea distributions we use $\alpha'_g=0.15
\mbox{GeV}^{-2}$ and
\begin{equation}
b_{g(sea)}=2.58 \mbox{GeV}^{-2}+ 0.25 \mbox{GeV}^{-2} \ln
\frac{m^2}{Q^2+m^2}
\end{equation}
which describe well the slope parameter observed experimentally.
The corresponding parameters for other contributions can be found
in \cite{gk06}.

Estimations for the light meson production amplitudes are carried
out using $f_{\rho L}=0.209\gev$, $a_{\rho L}=0.75\gev^{-1}$;
$f_{\phi L}=0.221\gev$; $a_{\phi L}=0.7\gev^{-1}$.

\begin{figure}[h!]
\begin{center}
\begin{tabular}{cc}
\mbox{\epsfig{figure=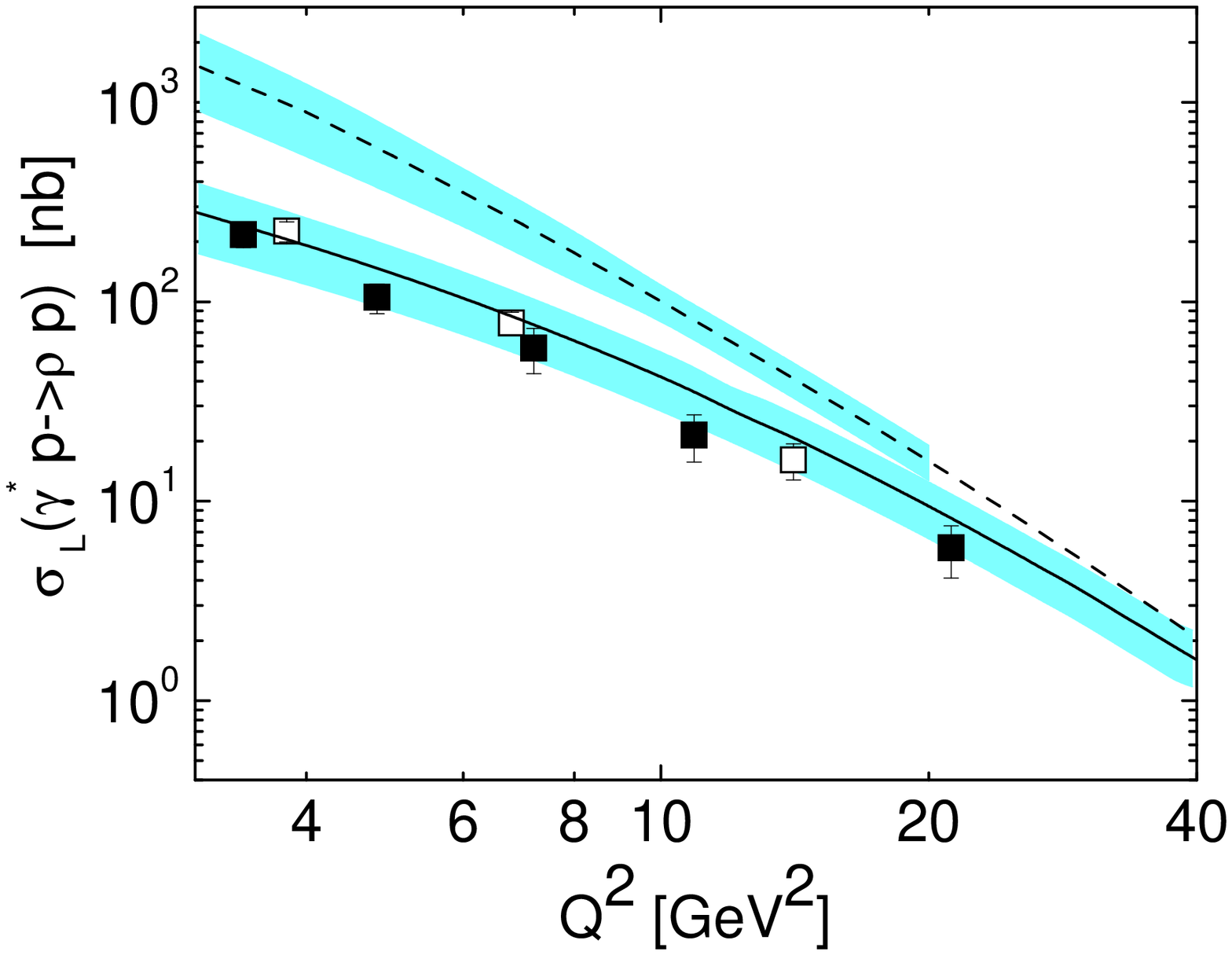,width=6.1cm,height=4.9cm}}&
\mbox{\epsfig{figure=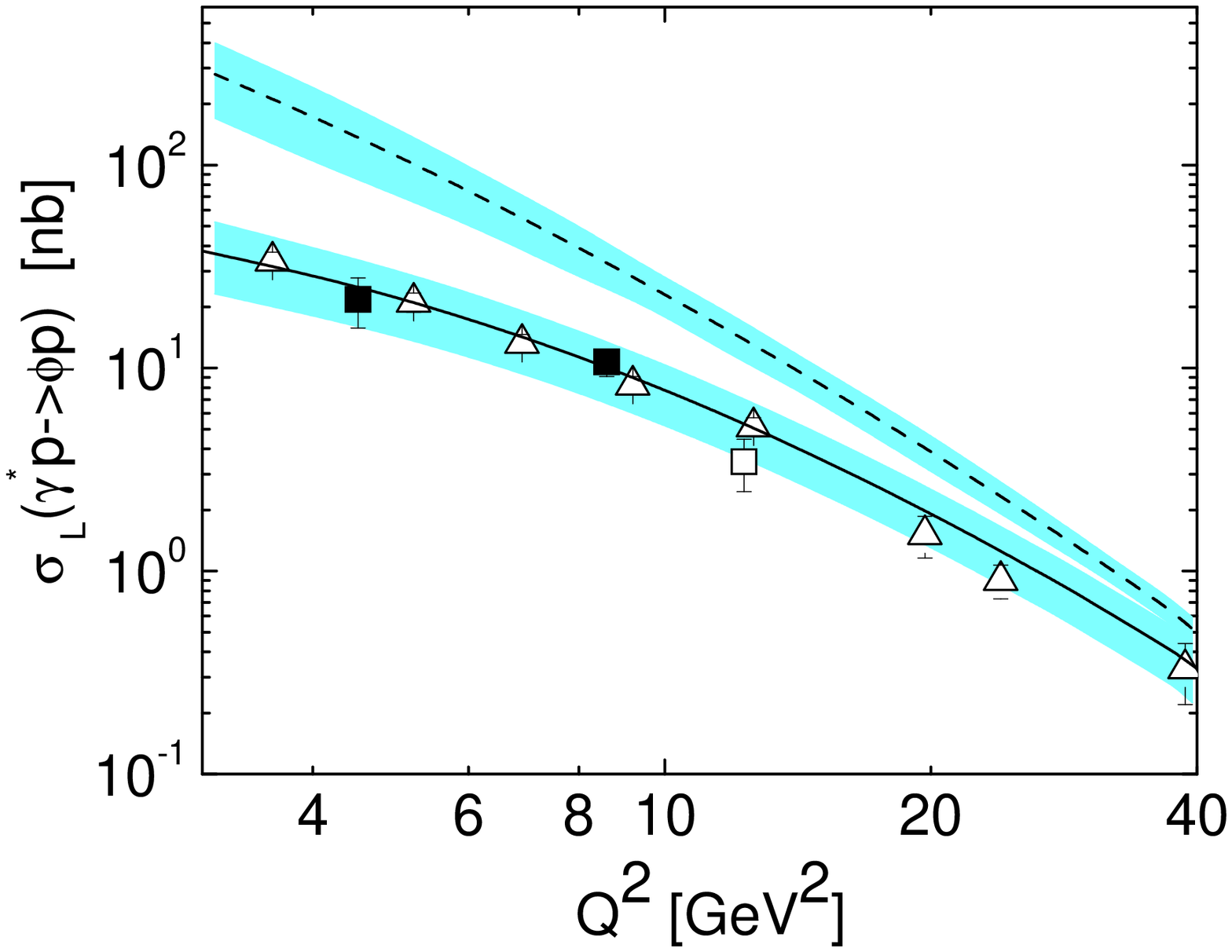,width=6.1cm,height=4.9cm}}\\
{\bf(a)}& {\bf(b)}
\end{tabular}
\end{center}
\caption{  {\bf(a)} Longitudinal cross sections of $\rho$
production at $W=75 \mbox{GeV}$.  {\bf(b)} Longitudinal cross
sections of $\phi$ production at $W=75 \mbox{GeV}$ with error band
from CTEQ6 PDFs uncertainties. Data are from H1 \cite{h1} -solid
symbols and ZEUS \cite{zeus} -open symbols.}
\end{figure}

\begin{figure}[h!]
\begin{center}
\begin{tabular}{cc}
\mbox{\epsfig{figure=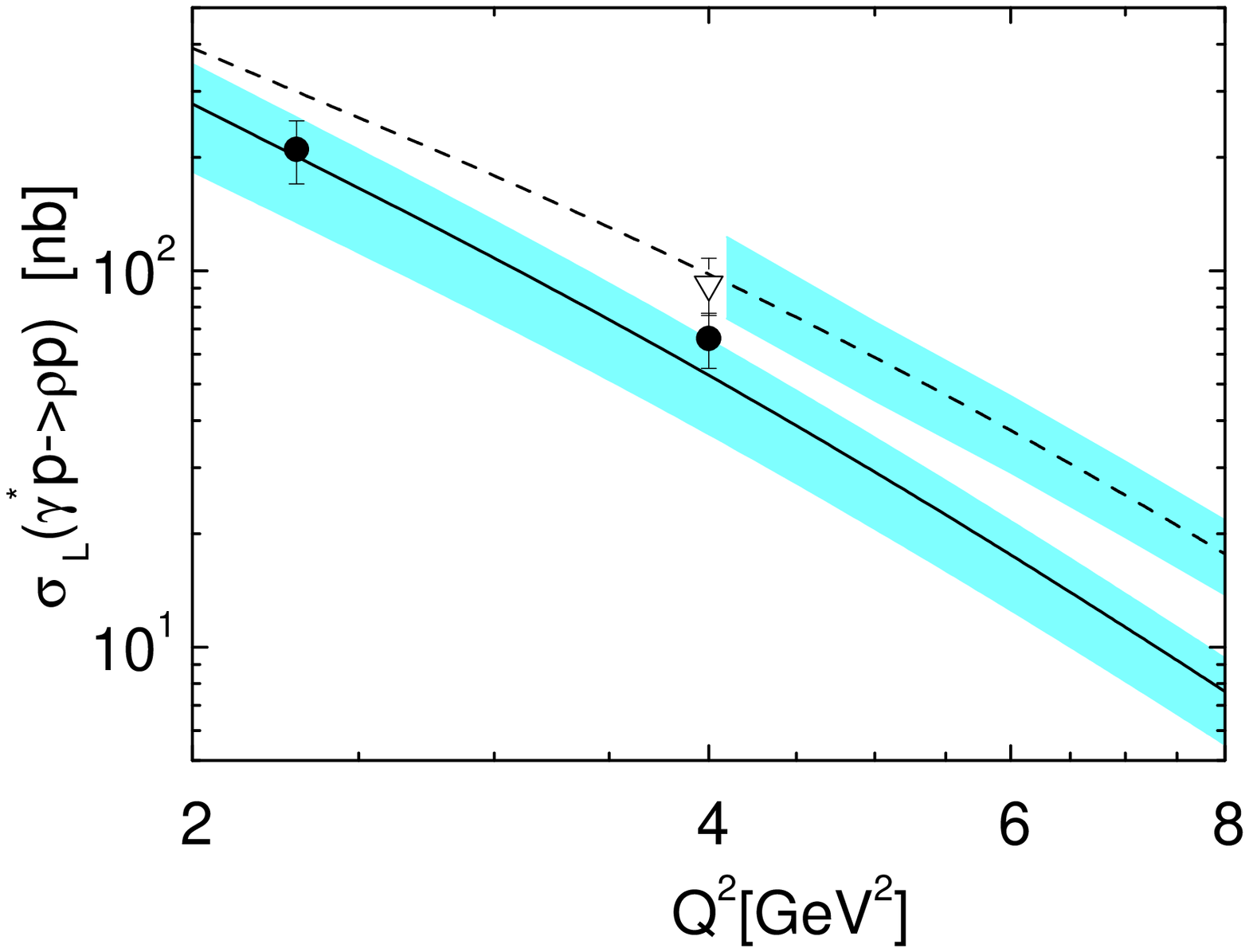,width=6.1cm,height=4.9cm}}&
\mbox{\epsfig{figure=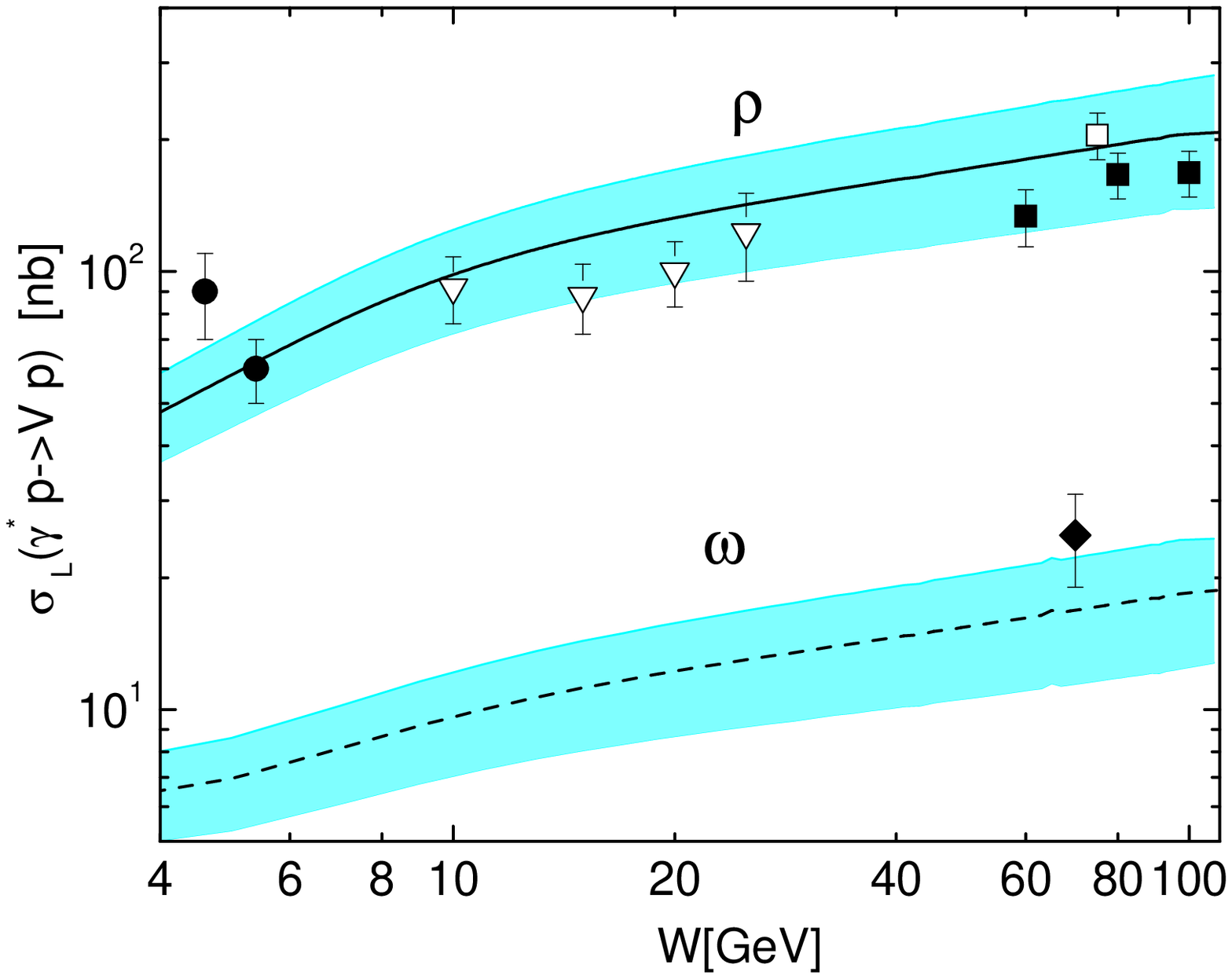,width=6.1cm,height=4.9cm}}\\
{\bf(a)}& {\bf(b)}
\end{tabular}
\end{center}
\caption{ {\bf(a)} Longitudinal cross sections of $\rho$
production at $W=5 \mbox{GeV}$ -full line and  $W=10 \mbox{GeV}$
-dashed line. {\bf(b)} Energy dependencies of cross section at
$Q^2=4 \mbox{GeV}^2$. Full line- longitudinal cross section of
$\rho$ production, dashed line- of $\omega$ production. Data are
from H1 \cite{h1}- solid squares, ZEUS \cite{zeus} - open square
and solid diamond \cite{zeusom}, E665 \cite{e665}- open triangles,
HERMES \cite{hermes} - solid circles, }
\end{figure}

At HERA energies we consider gluon and sea contributions. In this
energy range,  the valence quark effects can be neglected. The
cross section for the $\gamma^* p \to \rho p$ production
integrated over $t$ is shown in Fig. 2a. Good agreement with  H1
and ZEUS data \cite{h1,zeus} is found. The typical contribution of
the gluon--sea interference to $\sigma_L$ does not exceed 45\%
with respect to the gluon one. It is of the order of magnitude of
the uncertainties in the cross section (about 25-35 \%) from the
gluon GPD. The model results for the  $\phi$ production cross
section shown in Fig. 2b are consistent with the experiment
\cite{h1,zeus}.   The contribution of the gluon-sea quark
interference to the cross section of the $\phi$ production does
not exceed 25\%.

The leading twist results, which do not take into account effects
of a transverse quark motion, are presented in Fig. 2 too. One can
see that the $k_\perp^2/Q^2$ corrections in the hard amplitude
 are extremely important at low $Q^2$. They decrease
the cross section by a factor of about 10 at $Q^2 \sim
3\mbox{GeV}^2$.

In Fig.3a, the cross section for the $\rho$ production at HERMES
($W=5 \mbox{GeV}$) and our prediction for COMPASS ($W=10
\mbox{GeV}$) are presented. It can be seen that we describe
properly the available HERMES data \cite{hermes} and E665 data
 at $W=10 \mbox{GeV}$ \cite{e665}. In Fig 3.b, we show the
energy dependence of the cross section for the light meson
production. At high energies $W>10 \mbox{GeV}$, its energy
dependence is controlled by the gluon + sea contributions. At
lower energies valence quarks are essential. For the $\rho$
production at  HERMES $W =5\mbox{GeV}$ the valence quark
contribution to the cross section is about   40\%. For the case of
$\omega$ production \cite{zeusom} the quark contribution in this
energy range is about 65\%. This results in a more smooth energy
behavior of the $\omega$ cross section with respect to the $\rho$
one at $W \sim 5 \mbox{GeV}$, Fig. 3b.

The TT amplitude, which is essential for spin observables, is
calculated here at high energies where we study the gluon
contribution only. The analyses of quark effects in the TT
amplitude can be found in \cite{gk07t}. We  compare our results
for spin observables with experimental data at HERA energies where
the valence quarks are not essential. The results for the ratio of
the cross section with longitudinal and transverse photon
polarization
\begin{equation}\label{R}
R=\frac{\sigma_L}{\sigma_T}
\end{equation}
 for the $\rho$ and $\phi$ production are shown in Fig.4. We
describe properly available data form H1 and ZEUS experiments
\cite{h1,zeus}.\\[4mm]

\begin{figure}[h!]
\begin{center}
\begin{tabular}{cc}
\mbox{\epsfig{figure=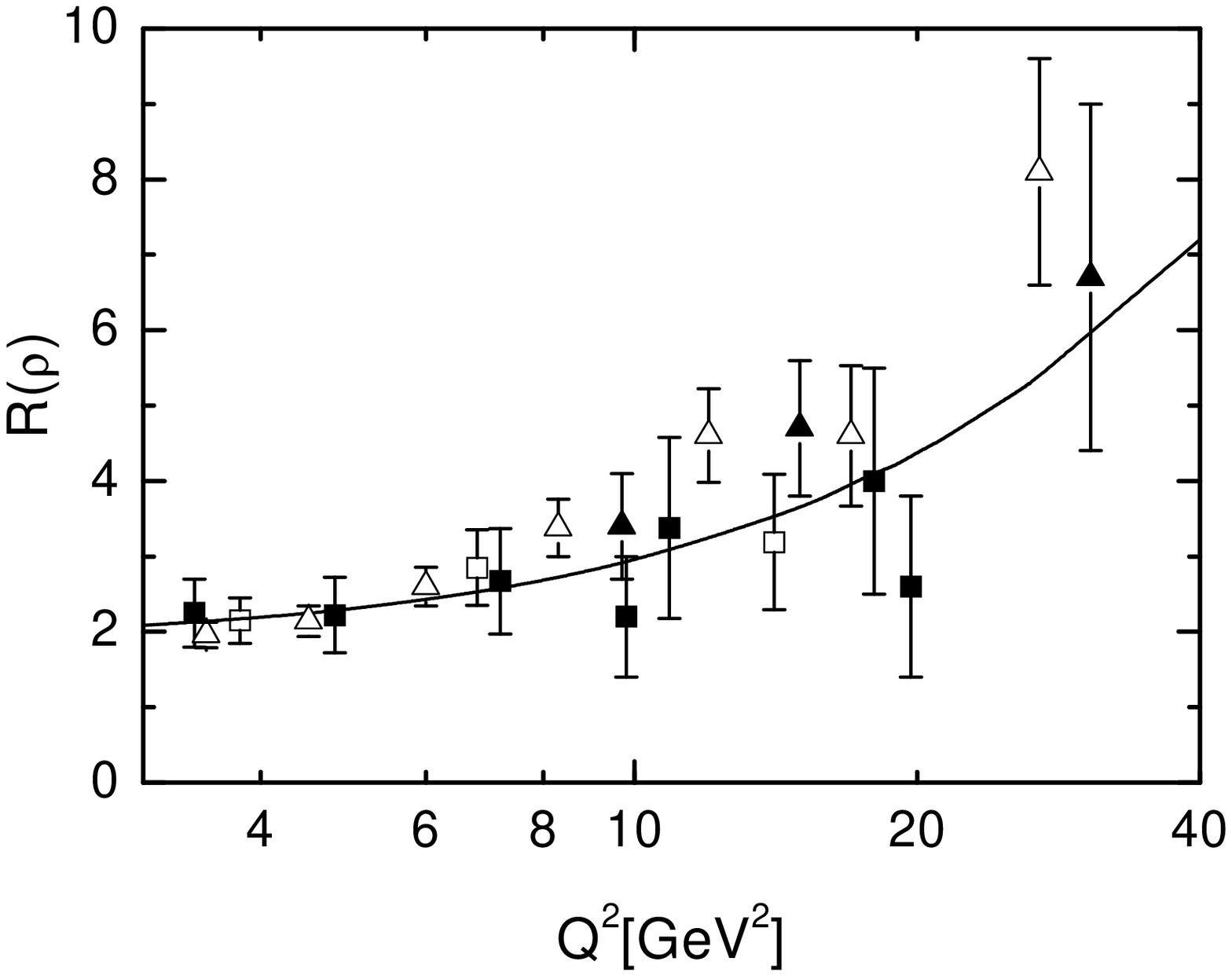,width=6.1cm,height=5.3cm}}&
\mbox{\epsfig{figure=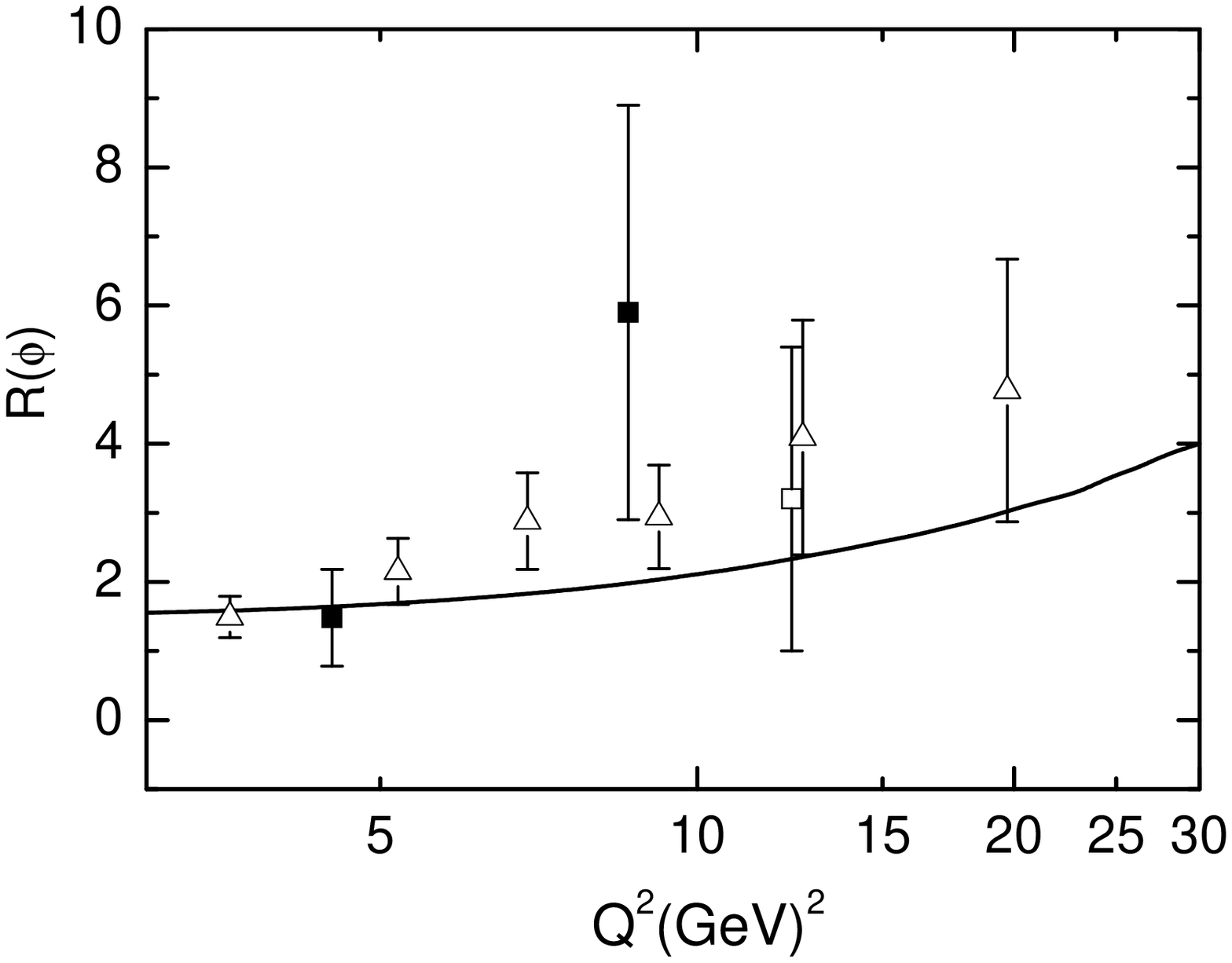,width=6.1cm,height=5.3cm}}\\
{\bf(a)}& {\bf(b)}
\end{tabular}
\end{center}
\caption{ {\bf(a)} $R$ ratio of $\rho$ production at $W=75
\mbox{GeV}$. {\bf(b)} $R$ ratio of $\phi$ production at $W=75
\mbox{GeV}$. Data are from H1 and ZEUS. }
\end{figure}

There are a few spin density matrix elements which are sensitive
to the LL and TT amplitudes. Three SDME are determined in terms of
the$R$ -ratio. In our approximation, we have
\begin{equation}
1-r^{04}_{00}=2\,r^1_{1-1}=-2\mbox{Im}
\,r^2_{1-1}=\frac{1}{1+\epsilon R}.
\end{equation}
Two  SDME are expressed in terms of the LL and TT amplitudes
interference
\begin{equation}
\mbox{Re} \,r^5_{10}=-\mbox{Im} \,r^6_{10} \propto
\mbox{Re}(M_{TT}\,M_{LL}^*).
\end{equation}
These SDME are relevant to the phase shift between these
amplitudes $\delta_{LT}$. In Fig.5, we present our results for the
SDME in the DESY energy range. Description of experimental data is
reasonable. Note that our model gives small  $\delta_{LT} \sim
2-3^o$ \cite{gk05}.

\begin{figure}[h!]
\begin{center}
\begin{tabular}{ccc}
\mbox{\epsfig{figure=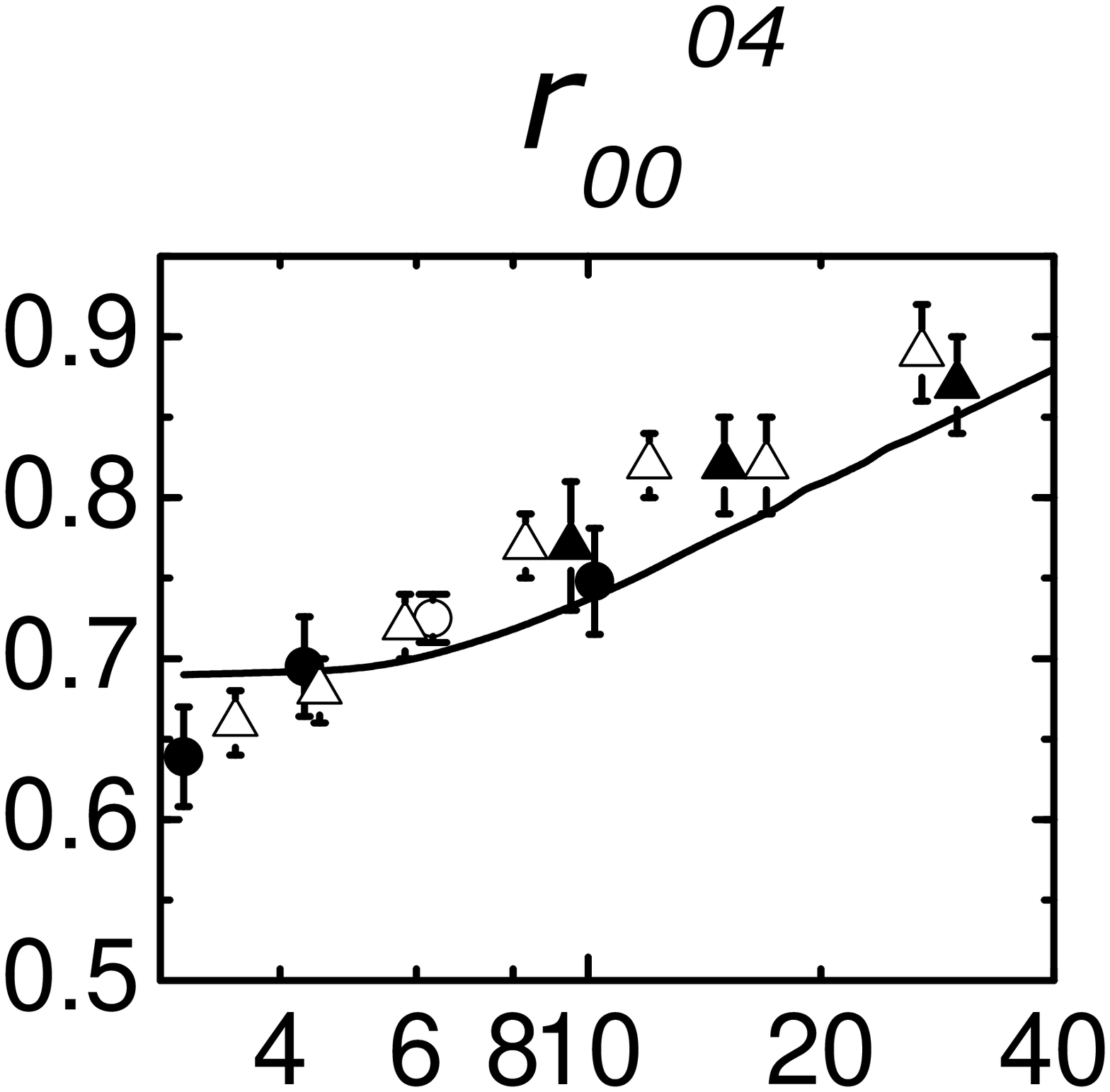,width=3.9cm,height=3.7cm}}&
\mbox{\epsfig{figure=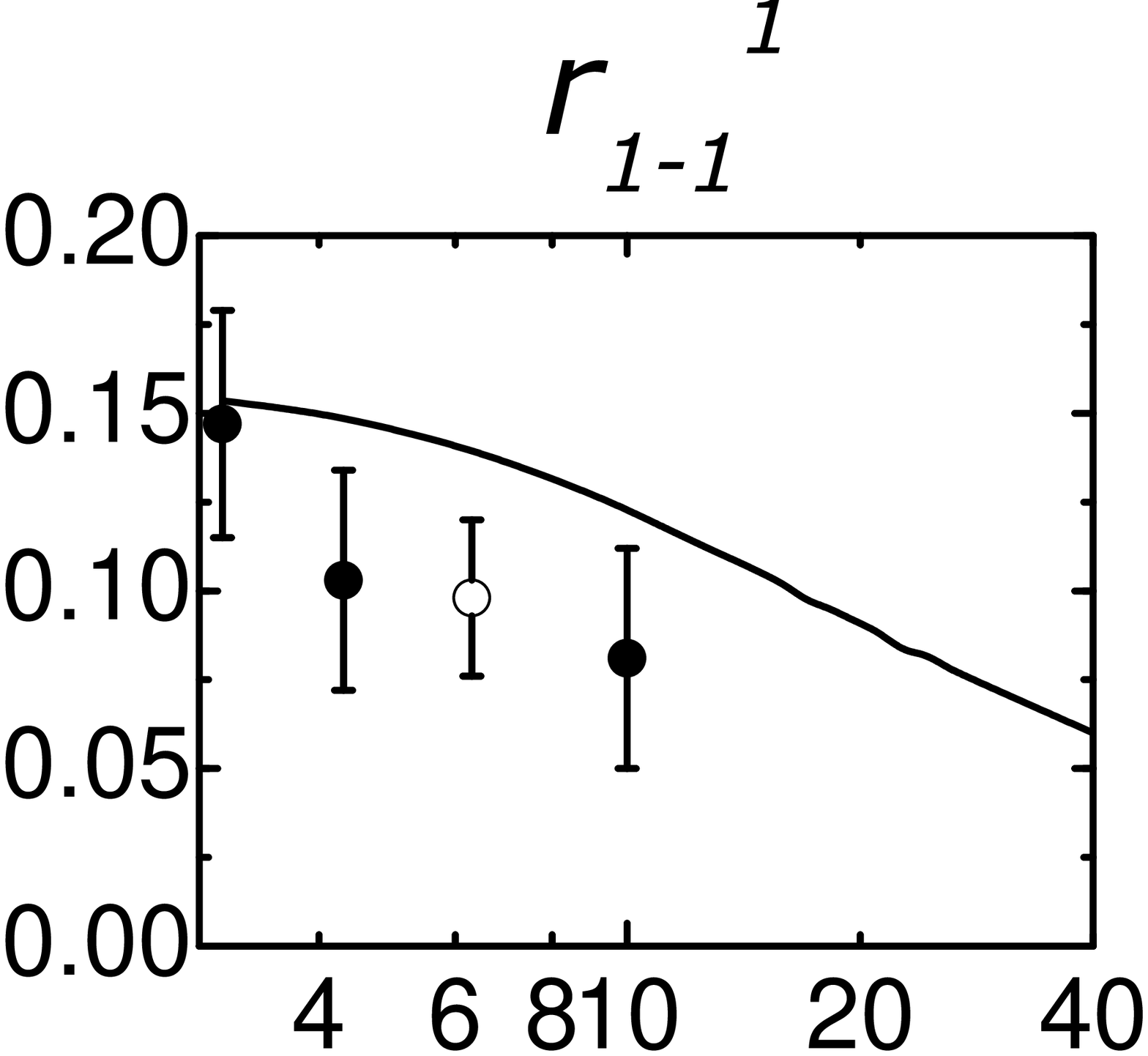,width=3.9cm,height=3.7cm}}&
\mbox{\epsfig{figure=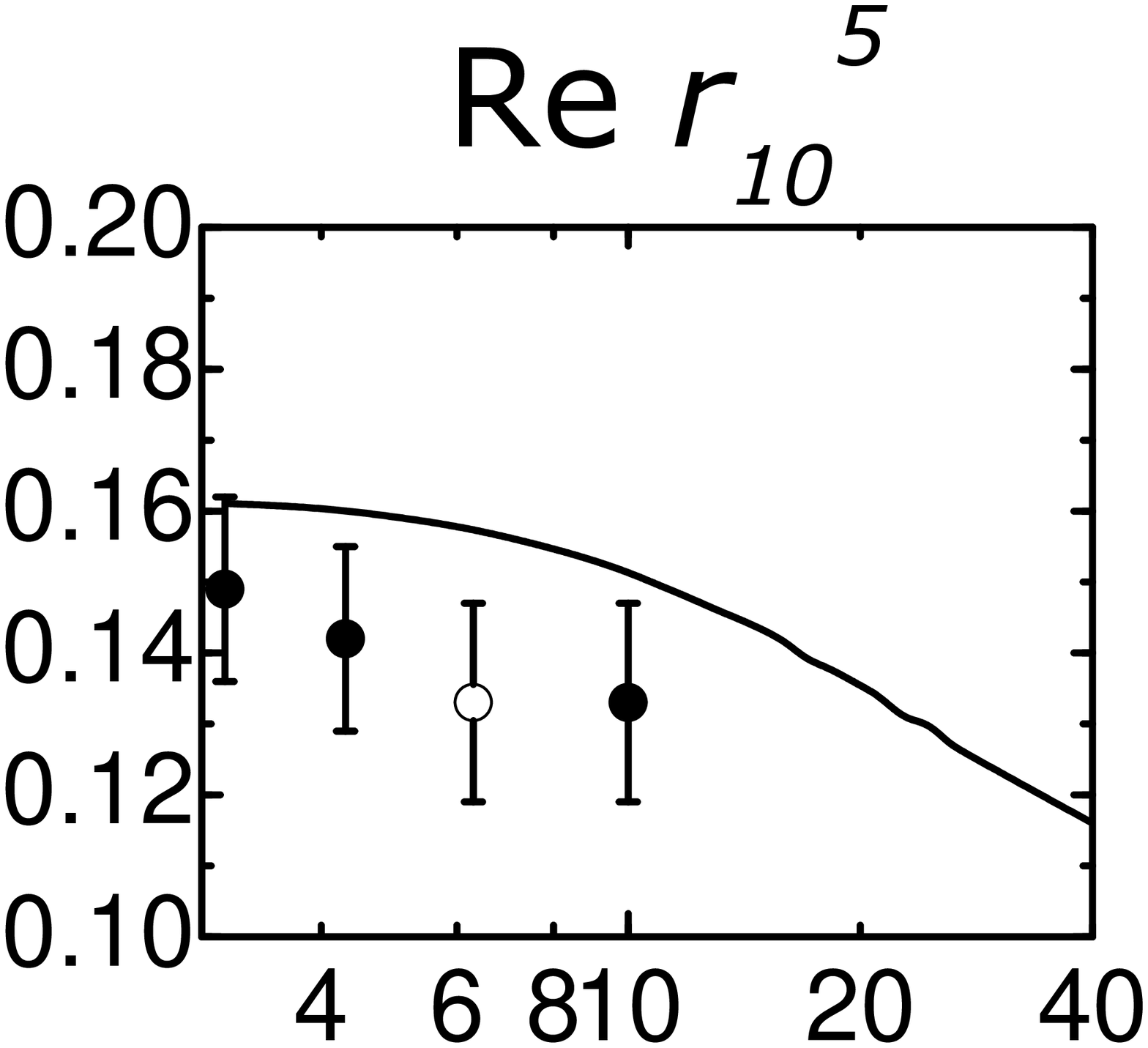,width=3.9cm,height=3.7cm}}\\
\end{tabular}
\phantom{aa}\vspace{-1mm} \centerline{$Q^2[\mbox{GeV}]^2$}
\end{center}
\caption{The $Q^2$ dependence of SDME on the $\rho$ production at
$t=-.15\gev^2$ and $W=75\gev$.   Data are taken from H1 and ZEUS.
}
\end{figure}

 \section{Conclusion or Summary}

Light vector meson electroproduction at small $x$ was analyzed
within the handbag model where the amplitude factorizes into a
hard subprocess  and   GPDs. The transverse quark momenta which
were considered in the hard propagators and in the vector meson
wave function regularize the end-point singularities in the
amplitudes with transversally polarized photons. This give a
possibility to calculate the TT amplitude and study spin effects
in the vector meson production in our model. The $k_\perp^2/Q^2$
corrections in the propagators decrease the cross section by a
factor of about 10 at $Q^2 \sim 3\mbox{GeV}^2$. As a result, we
describe  the cross section at $Q^2 \geq 3 \mbox{GeV}^2$.

In the model, a good description of the cross section from HERMES
to HERA energies \cite{gk06} is observed. It is found that the
gluon and sea contributions control the amplitude behaviour at
energies $W \geq 10 \mbox{GeV}$. Valence quarks are essential only
at HERMES energies, where their contribution to $\rho(\omega)$
cross section is about 40(65\%).

The model describes properly spin effects including the $R$ ratio
and SDME for the light meson production at HERA energies
\cite{gk05}.
 We would like to point out that study of SDME gives  important
information on different  $\gamma \to V$ hard amplitudes. The data
show a quite large phase difference $\delta_{LT} \sim 20-30^o$
\cite{hermes}.
 Our model with asmall phase difference
$\delta_{LT}$ gives a reasonable description of experimental data
at HERA. Unfortunately, the data on spin observables have large
experimental errors. To make clear  this and other  problems, new
experimental results on SDME are extremely important.

Thus, we can conclude that the vector meson photoproduction at
small $x$ is an excellent tool  to probe gluon and quark GPDs.

 \bigskip

 This work is supported  in part by the Russian Foundation for
Basic Research, Grant 06-02-16215  and by the Heisenberg-Landau
program.


\bigskip

 \begin{thebibliography}{99}
\bibitem{gk05} S.V. Goloskokov, P. Kroll,
  Euro. Phys. J. C{\bf 50}, (2007) 829.
\bibitem{gk06} S.V. Goloskokov, P. Kroll,
  Euro. Phys. J. C{\bf 42}, (2005) 281.
\bibitem{gk07t} S.V. Goloskokov, P. Kroll, arXiv:0708.3569 [hep-ph].
\bibitem{fact} X.\ Ji, Phys. Rev. D{\bf 55}, (1997) 7114;\\
A.V.\ Radyushkin,  Phys. Lett. B{\bf 380}, (1996)  417;\\
J.C.\ Collins, {\it et al.}, Phys.\ Rev. D{\bf 56}, (1997) 2982.
\bibitem{mpw} L.\ Mankiewicz, G.\ Piller and T.\ Weigl,   Eur.\ Phys.\ J.
C{\bf  5}, (1998)  119.
\bibitem{mp} L.\ Mankiewicz and G.\ Piller,
   Phys.\ Rev. D{\bf 61}, (2000) 074013;\\
I.V.\ Anikin and O.V.\ Teryaev, Phys.\ Lett. B{\bf 554}, (2003)
51.
\bibitem{sterman} J.\ Botts and G.\ Sterman,
 Nucl.\ Phys. B{\bf 325}, (1989) 62.
\bibitem{h1} C.~Adloff  et al.  [H1 Collaboration],
                        Eur.\ Phys.\ J. C{\bf 13}, (2000) 371;\\
                        C.~Adloff  et al.
                        [H1 Collaboration], Phys.\ Lett. B{\bf
                        483}, (2000)  360.
\bibitem{zeus} J.~Breitweg et al. [ZEUS Collaboration],
  Eur.\ Phys.\ J. C{\bf 6}, (1999) 603;\\
 S. Chekanov et al.  [ZEUS Collaboration], Nucl.Phys. B{\bf 718}, (2005) 3.
\bibitem{hermes} A.~Airapetian {\it et al} [HERMES collaboration],
     Eur.\ Phys.\ J.\ C{\bf 17}, (2000) 389;\\
      A. Borissov, [HERMES collaboration], "Proc. of Diffraction 06", PoS
        (DIFF2006), 014.
\bibitem{e665}  M.~R.~Adams {\it et al.}  [E665 Collaboration],
    Z.\ Phys.\  C {\bf 74}, (1997) 237.
\bibitem{koerner} J.\ Bolz, J.G.\ K\"orner and P.\ Kroll,
 Z.\ Phys. A{\bf  350},(1994) 145.
\bibitem{mus99} I.~V.~Musatov and A.~V.~Radyushkin,
  Phys.\ Rev. D{\bf 61}, (2000) 074027.
\bibitem{CTEQ} J.~Pumplin,et al.,
 JHEP {\bf 0207}, (2002) 012.
\bibitem{vinn} A.~V.~Vinnikov,
  hep-ph/0604248.

\bibitem{zeusom} J.~Breitweg et al. [ZEUS Collaboration],
Phys.\ Lett. B{\bf 487}, (2000) 273.
 \end{thebibliography}
\end{document}